\documentclass[twocolumn,prl,showpacs]{revtex4}
\usepackage{psfrag}
\usepackage{graphicx}
\usepackage{amsmath}
\usepackage{amssymb}

\begin{document}
\title{Four-atom period in the conductance of monatomic Al wires}
\author{K. S. Thygesen}
\affiliation{Center for Atomic-scale Materials Physics, \\
Department of Physics, Technical University of Denmark, DK - 2800 Kgs. Lyngby, Denmark}
\author{K. W. Jacobsen}
\affiliation{Center for Atomic-scale Materials Physics, \\
Department of Physics, Technical University of Denmark, DK - 2800 Kgs. Lyngby, Denmark}

\date{\today}

\begin{abstract}
We present first principles calculations based on density functional
theory for the conductance of monatomic Al wires between Al(111) electrodes. In
contrast to the even-odd oscillations observed in other metallic
wires, the conductance of the Al wires is found to oscillate with a
period of 4 atoms as the length of the wire is varied. Although local
charge neutrality can account for the observed period it leads to an
incorrect phase. We explain the conductance behavior using a resonant
transport model based on the electronic structure of the infinite wire.

\end{abstract}
\maketitle

Metallic chains of single atoms represent the ultimate limit of the
miniaturization of electrical conductors. Apart from possible
technological applications as interconnects in molecular electronics,
the simple structure of these monatomic wires makes them an ideal test
ground for developing and validating our understanding of electron
transport on the nanometer length scale. The conductance of monatomic
wires has been investigated theoretically by several authors using a
number of different
techniques~\cite{lang97,lang98,sim01,kobayashi00,tsukamoto02,havu02}.
Lang was the first to study the dependence of the conductance on the
number of atoms in the wire. In 1997 he found that the conductance of
a chain of Na atoms between jellium electrodes oscillates with a
period of 2 atoms as the length of the chain is varied~\cite{lang97},
and the following year Lang and Avouris observed the same behavior for
chains of C atoms~\cite{lang98}. This even-odd effect was also found
in Na wires by Tsukamoto and Hirose~\cite{tsukamoto02} using a method
similar to that of Lang, and by Sim et al.~\cite{sim01} who combined
cluster density functional theory (DFT) calculations with the Friedel
sum rule to obtain the conductance.  Recently, the even-odd
oscillations have been confirmed experimentally for Au, Pt and Ir by
means of the mechanically controlled break junction~\cite{smit03}.

The fact that a large number of metals of rather distinct character
exhibit the even-odd effect has lead to the suggestion that this is a
universal feature of atomic wires~\cite{smit03}.  In this paper we
present calculations based on DFT showing that the conductance of
monatomic Al wires varies in an oscillatory manner as the number of
atoms in the wire is changed. In contrast to the even-odd effect,
however, the oscillation has a period of 4 atoms. Based on a resonant
transport picture we show that local charge neutrality can account for
the observed 4-atom period. For long wires, however, we find that the
occupancy of molecular resonant states and thus the net charge located
on the wire is determined by the position of the discrete energy
spectrum of the free wire relative to the Fermi level of the
electrodes. In general these two effects compete in determining the
phase of the conductance oscillation. In the case of Al we find that
the latter effect dominates for wires containing more than about 4
atoms, whereas charge neutrality is more important for shorter wires.
By relating the electronic structure of the free wire to that of the
infinite wire, we arrive at a simple picture for the position of the
free-wire valence resonances which accounts for the conductance both
qualitatively and quantitatively. In this picture the filling factor
of the infinite wire determines the period of the conductance
oscillation: a half filled band implies the even-odd effect while a
filling factor of 0.25 leads to a 4-atom period.

\begin{figure}
\begin{center}
\includegraphics[width=0.9\linewidth,bb=31 172 464 322, clip]{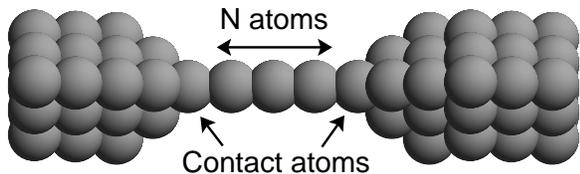}
\end{center}
\caption[system]{\label{fig1}The $N=3$ wire connecting Al(111)
  electrodes via a 3-atom basis. This structure constitutes the supercell in
  the DFT calculations and forms the scattering region in the
  conductance calculations.}
\end{figure}

We represent the combined electrode-wire system in a supercell
containing the wire itself together with seven 3x3 sections of the
(111) planes in bulk aluminum. An example of such a supercell is shown
in Fig.~\ref{fig1}. Periodic boundary conditions are imposed on the
supercell in the directions perpendicular to the wire axis, and for
the DFT calculations also in the parallel direction. The calculational
procedure consists of two main steps: (i) first a realistic wire
configuration is found by relaxing the ion positions as well as the
electrode-electrode distance as determined by the length of the
supercell. During the relaxation the wire is constrained to be linear
and ions below the surface layer are fixed in the bulk crystal
structure. These simulations were performed within DFT using a
plane-wave based pseudopotential code~\cite{dacapo} with an energy
cutoff of 15 Ry for the plane-wave expansion. The ion cores are
replaced by ultrasoft pseudopotentials~\cite{vanderbilt90} and to
treat exchange and correlation we use the PW91 functional~\cite{pw91}.
The bond lengths in the wires lie in the range 2.36~{\AA} to
2.50~{\AA} with the shorter bond lengths achieved closer to the center
of the wire. For the longer wires the bond lengths vary less and
approach a value of 2.39~{\AA} which is also found for the infinite
wire. No dimerization is found in accordance with previous results for
infinite linear chains of Al atoms~\cite{rubio96}. We mention that
Ayuela et al. have found that an infinite linear Al chain has a
nonmagnetic groundstate for interatomic distances smaller than
2.7~{\AA}~\cite{ayuela02}.  (ii) The self-consistent effective
potential generated by the DFT code is used as input to the Green's
function transport code described in detail in Ref.~\cite{thygesen03}.
The effective potential of the supercell forms the scattering region
in the transport problem while the potential in the semi-infinite
leads is obtained from a DFT bulk calculation. In this way the
scattering region and the leads are treated on an equal footing taking
their full atomic and electronic structure into account. The
conductance is calculated from the Green's function of the scattering
region which is represented in terms of a system-independent basis set
containing wavelets of compact support. All conductance calculations
have been converged with respect to the size of the basis set. Due to
the limited size of the supercell in the directions perpendicular to
the wire, the two-dimensional Brillouin zone is sampled by 8 special
$\mathbf k$-points, both in the transport and the DFT calculations.

\begin{figure}
\begin{center}
\includegraphics[width=0.9\linewidth,bb=0 13 707 524,clip]{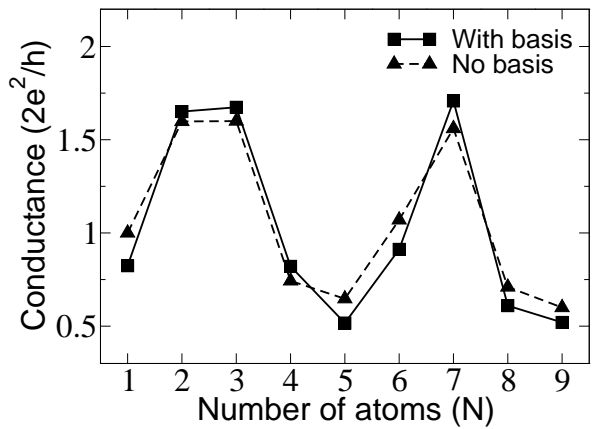}
\end{center}
\caption{\label{fig2}Calculated conductance as a
  function of wire length with and without a 3-atom basis as contact
  to the electrodes.}
\end{figure}

In Fig.~\ref{fig2} we have plotted the calculated conductance of
Al wires containing 1 to 9 atoms and attached to the (111) surface
either directly or via a 3-atom basis. We regard an atom as
part of the wire if it has a coordination number of 2, see
Fig.~\ref{fig1}. With this convention the wire couples to
the electrodes via a single atom which we refer to as the contact
atom. The conductance oscillates with a period
of 4 atoms taking values in the range $0.5G_0$ to $1.7G_0$. The
oscillation amplitude increases approximately 10\% when the 3-atom
basis is used as contact to the electrodes, but otherwise
no difference is found between the two geometries. 

\begin{figure}
\begin{center}
\includegraphics[width=0.9\linewidth,bb=0 10 715 537,clip]{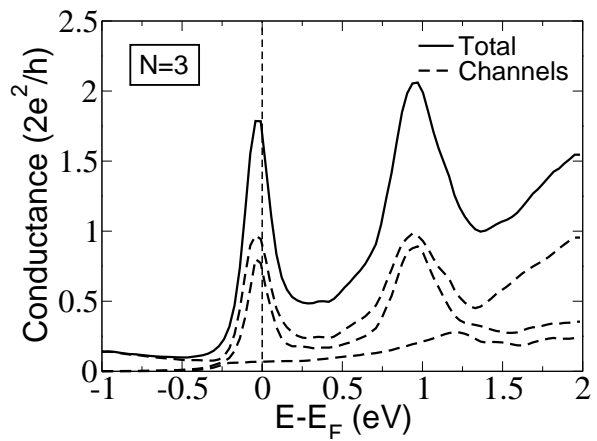}
\end{center}
\caption[channel]{\label{fig3}Eigen-channel resolved conductance for
  the $N=3$ wire shown in Fig. \ref{fig1}. The two similar
  $\pi$-channels together with the $\sigma$-channel are indicated by the
  dashed lines, while the solid line shows the total conductance.}
\end{figure}

We can gain some insight into the
electronic structure of the current carrying states by studying the
eigen-channel resolved conductance~\cite{brandbyge} as shown in Fig.~\ref{fig3} for $N=3$. The valence configuration of atomic Al is $3s^2
3p$, which for the infinite wire leads to a fully occupied $\sigma$-band
and a degenerate $\pi$-band with filling factor $f=0.25$. 
Three channels contribute to the
total conductance: two similar $\pi$-channels and a single
$\sigma$-channel. The two $\pi$-channels constitute the main part of the
conductance, while
the $\sigma$-channel is a tunneling channel whose contribution decays
exponentially with the wire length from $\sim 0.2G_0$ at $N=1$
to $\sim 10^{-4}G_0$ at $N=9$. 

To understand the origin of the conductance oscillations we first
relate the conductance to the electronic structure of the wire. When a
free $N$-atom wire is connected to bulk electrodes, the molecular
orbitals on the wire hybridize with extended states in the electrode.
As a result of this mixing, the $n$'th energy level of the free wire
is broadened into a resonance described by the projected density of
states~\cite{newns69}
\begin{equation}
\rho_n(\varepsilon)=\frac{\pi^{-1}\Delta_n(\varepsilon)}{(\varepsilon-\varepsilon_n-\Lambda_n(\varepsilon))^2+\Delta^2_n(\varepsilon)}.
\end{equation}
For each molecular level, the coupling to the electrodes defines a unique
orbital, the group orbital, which is located on the contact atoms. In terms of the group orbital we can
write $\Delta_n(\varepsilon)=\pi |V_n|^2\rho_{c,n}(\varepsilon)$, where $V_n$ is the
matrix element coupling the $n$'th molecular level with its
corresponding group orbital and $\rho_{c,n}$ is the projected density of states for the group orbital in the
absence of the wire. For simplicity we shall not distinguish between
the group orbitals and thus we set $\rho_{c,n}=\rho_c$. 
If we furthermore neglect interference
between the different molecular levels we obtain a simple expression for the conductance 
\begin{equation}\label{eq:conductance}
G=\frac{2e^2}{h}\sum_{n}\pi^2 |V_n|^2\rho_c(\varepsilon_F) \rho_n(\varepsilon_F),
\end{equation}
where the sum extends over all molecular levels.
From this expression we expect the position of the resonances relative
to the Fermi level of the electrodes to be a crucial parameter for the conductance.

So far, local charge neutrality has been considered the main cause of
the even-odd oscillations in monovalent
wires~\cite{sim01,havu02,smit03}. Indeed, for monovalent wires
consisting of an odd number of atoms, local charge neutrality pins the
Fermi level to the middle of a resonance resulting in a high
conductance. On the other hand, for even-numbered wires the Fermi
level lies off resonance leading to a lower conductance. In the case
of Al the degeneracy of the $\pi$-orbitals on the free wire and the
fact that each Al atom provides one electron to the $\pi$-system,
implies that local charge neutrality would cause oscillations in the
conductance with a period of 4 atoms. However, the resulting phase of
the oscillation would be incorrect. The maxima (minima) occur for
half-filled (filled) resonances which should correspond to
$N=2,6,\ldots$ ($N=4,8,\ldots$). These numbers are off by 1 atom as
compared to Fig.~\ref{fig2}. Consequently, charge neutrality alone
cannot account for the conductance of the Al wires. At this point it
is interesting to notice that different results have been obtained for
the phase of the conductance oscillation in Na wires. As opposed to
Tsukamoto and Hirose and Sim et al., Lang~\cite{lang97} has found a
higher conductance for even-numbered than for odd-numbered Na chains.
Calculations by Havu et al.~\cite{havu02} for Na wires between jellium
cones show that it is possible to change the phase of the conductance
oscillation from even-odd to odd-even by varying the lead opening
angle.

\begin{figure}
\begin{center}
\includegraphics[width=0.8\linewidth,bb=0 180 556 676,clip]{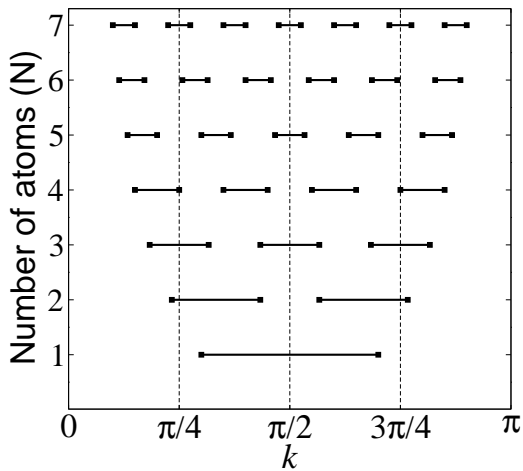}
\end{center}
\caption{\label{fig4}Evolution in $k$-space of the discrete spectrum of
  a free $N$-atom wire. The interatomic distance in the
  wire is set to 1. The horizontal lines symbolize the
  width of the corresponding resonance formed when the free wire is
  coupled to electrodes.}
\end{figure}

To explain the behavior of the conductance, we need to examine more
closely the position of the resonances relative to the Fermi level of
the electrode-wire-electrode system. 
As a first step in this
direction we consider a linear chain of $N$ sites coupled by nearest
neighbor hopping. The Hamiltonian is
\begin{equation}\label{eq:modelham}
H=\sum_{i=1}^{N} \varepsilon_0 c^{\dagger}_i c_i +
\sum_{i=1}^{N-1}t(c^{\dagger}_{i+1}c_i+c^{\dagger}_{i}c_{i+1})
\end{equation}  
where $c_i$ annihilates an electron at site $i$. The eigenvalues, $\varepsilon_n^N$, of
$H$ can be expressed in terms of the band
structure of the infinite chain of coupled sites, $\varepsilon(k)$, as follows:
\begin{equation}\label{eq:eigenvalue}
\varepsilon^N_n=\varepsilon(\frac{n\pi}{N+1})\:,\:n=1,\ldots,N
\end{equation}
where we have set the inter-site distance to 1. Although the above
relation is only strictly valid for the model Hamiltonian
(\ref{eq:modelham}), we can still use it to estimate the spectrum of a
free $N$-atom wire in terms of the band structure of the infinite
wire. As already described, the coupling of the free wire with the
electrodes converts the discrete energy spectrum into a set of
resonances. Assuming $\rho_c(\varepsilon)$ to be constant the $n$'th
resonance of the $N$-atom wire becomes a Lorentzian of width
$\Delta_{N,n}=\pi |V_{N,n}|^2\rho_c$. Since we are only interested in
resonances close to the Fermi energy, it is reasonable to assume an
energy independent coupling, $V_{N,n}=V_N$. Furthermore, the extended
nature of the eigenstates of the free wire causes $|V_N|^2 \propto
1/N$. In Fig.~\ref{fig4} we have illustrated in $k$-space the
evolution of the discrete spectrum of a free wire as a function of the
wire length. For each eigenvalue we have indicated the width of the
corresponding resonance. The eigenvalues are converted to energy via
the valence band of the infinite wire and relation
(\ref{eq:eigenvalue}). To find the position of the Fermi level in the
electrode-wire-electrode system we note that the Fermi energy is fixed
by the macroscopic electrodes, and thus independent of the length of
the wire. Since we do not expect a wire to charge up as it grows
longer the net charge per atom must vanish in the limit of a long
wire, and the band filling, $f$, must approach that of the infinite
wire. In Fig.~\ref{fig4} we have indicated the Fermi level for the Al
wire by a vertical line at $k=\pi/4$ corresponding to $f=0.25$.  As
$N$ increases the resonances move down through the Fermi level in a
systematic fashion. The coincidence of the Fermi level with the center
of a resonance is a periodic event of period 4 starting at $N=3$.
Like the local charge neutrality, this picture explains the observed
period of the conductance oscillation but in addition it provides
information about the actual position of the resonances. The first
maximum (minimum) is predicted to occur at $N=3$ ($N=5$), which indeed
gives the correct phase as compared to Fig.~\ref{fig2}.  The vertical
line at $k=\pi/2$ indicates the position of the Fermi level for metals
with a half-filled valence band such as Na, Au and C.  Clearly, our
model accounts for the even-odd behavior observed in these wires.

According to the filling of $\pi$-resonances as predicted by
Fig.~\ref{fig4}, the $N$-atom Al wire should hold approximately $N-1$
valence electrons. For short wires the electrostatic potential from
the missing electron drags the resonances down in energy thereby
increasing the occupancy. The resulting position of the
resonances must be determined self-consistently and will in general lead to an
occupation somewhere between $N-1$ and $N$. For short wires the simple picture
is therefore modified by a charge-neutralization process which tend to
shift down the phase of the conductance oscillation by one atom. This
effect explains the relatively high conductances found for $N=1,2$ as
compared to $N=5,6$.     
\begin{figure}
\begin{center}
\includegraphics[width=0.9\linewidth,bb=0 15 707 537,clip]{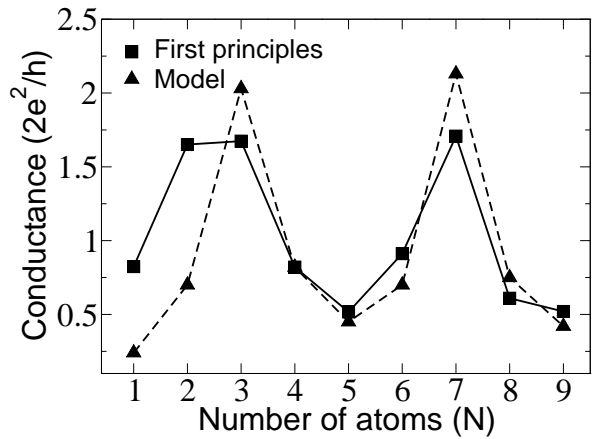}
\end{center}
\caption{\label{fig5}The conductance of Al wires as calculated from first
  principles (solid) and a simple model (dashed).}
\end{figure}

To test our simple model quantitatively we have used it to reproduce
the first principles results. We use Eq. (\ref{eq:conductance}) for
the conductance and assume a Lorentzian shape of the projected density
of states. The molecular levels given in Eq. (\ref{eq:eigenvalue}) are
obtained via the $\pi$-band of the infinite Al wire calculated within
DFT. From the first resonance in Fig.~\ref{fig3} we read off
$\Delta_{N=3} = 0.15$ eV as the half-width at half-maximum. Due to the
$1/N$ scaling, all resonance widths then follow from $\Delta_{N=3}$.
The result of the model is seen in Fig.~\ref{fig5} together with the
first principles calculation, repeated here for convenience.  Except
for the shortest wires ($N=1,2$) where charge transfer as well as
conduction through the $\sigma$-channel is important, the model gives
a good description of the conductance with an average deviation of
less than $15\%$. The fact that the model gives values above $2G_0$ is
a consequence of the neglected interference between different
molecular levels. Finally, we note that the $1/N$-dependence of the
resonance width together with the $N$-dependence of the resonance
spacing implies the existence of conductance oscillations also for
longer wires.  This is in agreement with previous results in
Ref.~\cite{zeng02}.

In summary, our calculations show that in contrast to the even-odd
effect the conductance of monatomic Al
wires oscillates with a 4-atom period as the number of atoms in the
wire is varied. This behavior is explained by combining a resonant
transport picture with the electronic structure of the free Al
wire. By relating the electronic structure of a free wire to the
valence band of the infinite wire, we find that the period of
oscillation is determined by the filling factor of the valence band of
the infinite wire. For long wires ($>4$ atoms) a simple model based on these ideas was shown to
reproduce the first principles conductances while for
short wires the simple picture is modified by charge transfer
and tunneling effects.

The Center for Atomic-scale Materials Physics is sponsored by the
Danish National Research Foundation. We acknowledge support from the
Danish Center for Scientific Computing through Grant No. HDW-1101-05.


\end{document}